\documentclass[reprint,showpacs,preprintnumbers,amsmath,amssymb,aip]{revtex4-1}
\usepackage{graphicx,wasysym}
\usepackage{dcolumn}
\usepackage{bm}
\usepackage{bbm}
\usepackage{natbib}
\usepackage{tabulary}
\usepackage{color}
\usepackage{hyperref}

\usepackage{color}

\usepackage{graphicx}
\usepackage{amsmath}
\usepackage{color}
\usepackage{float}
\usepackage{bm}

\newcommand{\kB}{k_\mathrm{B}}

\begin{document}

\title{Variational design principles for nonequilibrium colloidal assembly}

\author{Avishek Das}
\affiliation{Department of Chemistry, University of California, Berkeley CA 94609 \looseness=-1}
\author{David T. Limmer}
\email[Corresponding author: ]{dlimmer@berkeley.edu} 
\affiliation{Department of Chemistry, University of California, Berkeley CA 94609 \looseness=-1}
\affiliation{Kavli Energy NanoScience Institute, Berkeley, CA 94609 \looseness=-1}
\affiliation{Materials Science Division, Lawrence Berkeley National Laboratory, Berkeley, CA 94609 \looseness=-1}
\affiliation{Chemical Science Division, Lawrence Berkeley National Laboratory, Berkeley, CA 94609\looseness=-1}

\date{\today}
\begin{abstract}
Using large deviation theory and principles of stochastic optimal control, we show that rare molecular dynamics trajectories conditioned on assembling a specific target structure encode a set of interactions and external forces that lead to enhanced stability of that structure. Such a relationship can be formulated into a variational principle, for which we have developed an associated optimization algorithm and have used it to determine optimal forces for targeted self-assembly within nonequilibrium steady-states. We illustrate this perspective on inverse design in a model of colloidal cluster assembly within linear shear flow. We find that colloidal clusters can be assembled with high yield using specific short-range interactions of tunable complexity. Shear decreases the yields of rigid clusters, while small values of shear increase the yields of nonrigid clusters. The enhancement or suppression of the yield due to shear is rationalized with a generalized linear response theory. By studying 21 unique clusters made of 6, 7 or 8 particles, we uncover basic design principles for targeted assembly out of equilibrium. 
\end{abstract}

\pacs{}

\keywords{} 
\maketitle

\section{Introduction}
The self-assembly of soft and biological matter out of equilibrium can result in novel structures and dynamical responses not constrained by thermodynamic considerations.\cite{rabani2003drying,stenhammar2013continuum, chremos2010ultra,toner2005hydrodynamics,li2020non,arango2018understanding} 
The microscopic violation of detailed balance in such systems can be used to design a wide range of functional materials with enhanced thermomechanical, optoelectronic or drug-delivery properties.\cite{guan2018fabrication,tsipotan2016laser,chevreuil2018nonequilibrium} 
Predictive inverse design to drive the assembly of target dissipative  structures requires a dynamical description of the system.\cite{hagan2006dynamic,whitelam2015statistical,whitelam2009role,grandpre2020entropy,newton2015rotational} 
We have developed a variational algorithm to automate the discovery of inverse design principles for colloidal self assembly in a nonequilibrium steady-state in molecular dynamics simulations. The algorithm uses a variational principle arising from rare dynamical fluctuations of the system in a trajectory ensemble, and optimizes the yield of target clusters with statistically estimated explicit gradients in the design parameter space. We demonstrate the performance of this algorithm by obtaining optimal design principles for the self-assembly of DNA-labeled colloids\cite{hormoz2011design} driven out of equilibrium by a shear flow. We expect that the ability to uncover general optimal inverse design principles away from equilibrium will enable bottom-up synthesis of new materials and elucidate the processes encoding structure in biological contexts.\cite{vauthey2002molecular,zhang2002design,zwicker2017growth} 

Self-assembly of nanoscale building blocks is increasingly used to engineer functional materials with novel properties arising from their complex nanostructures. Colloidal systems offer a versatile paradigm for inverse design towards a desired target structure due to the independent tunability of shape, valency and assembly environment.\cite{boles2016self} In thermodynamic equilibrium, stabilizing a target structure amounts to lowering its free energy, which is typically achieved by increasing both the interaction strength and specificity. This principle has been exploited to achieve the self-assembly of a variety of clusters and superlattices from colloids and nanocrystals with crystal facets decorated with organic ligands or DNA.\cite{mirkin1996dna,talapin2009quasicrystalline,damasceno2012predictive} 
However, employing these principles in practice requires mitigating dynamical effects like slow coarsening and kinetic trapping.\cite{whitelam2009role,whitelam2015statistical} 
Optimal forces for self-assembly must achieve a trade-off between slow relaxation at high interaction strengths, and slow growth at high interaction specificity.\cite{klotsa2013controlling,fullerton2016optimising} 
Self-assembly of colloids and biomolecules in nonequilibrium steady-states provide a route to decouple kinetics from stability and mitigate this tradeoff. 
Directive self-assembly has been achieved by driving the system with a constant supply of chemical fuel, or by applying external fields.\cite{grzybowski2017dynamic,ragazzon2018energy,heinen2019programmable,grunwald2016exploiting}
However, the design of such systems must confront the continuous supply of energy necessary to prevent the system from relaxing to equilibrium. Existing computational methods to discover inverse design principles for nonequilibrium self assembly are limited due to the configurational probability not following the Boltzmann distribution and the corresponding variational structure afforded by the free energy no longer being valid under such dissipative conditions. 

Recent advances in the theoretical treatment of the stochastic thermodynamics of nonequilibrium steady-states have made possible a trajectory ensemble description of self-assembly, treating structure and dynamics on an equal statistical footing.\cite{derrida2007non,seifert2012stochastic} This has enabled basic principles governing assembly away from equilibrium to be formulated.\cite{nguyen2016design,kuznets2020dissipation} In this work we develop a perspective and accompanying numerical technique based on these insights. Rather than considering the probability of observing a state and tuning its associated free energy, we consider the likelihood that a trajectory forms a specific structure as quantified by a stochastic action, and how that action is changed by modifying the intermolecular and applied forces. We show that fluctuations around a nonequilibrium steady-state encode the susceptibility of a system to assemble, in a manner analogous to a fluctuation-dissipation relationship. Further,  optimal forces that assemble a target structure while minimizing the change to the stochastic action satisfy a variational principle.\cite{chetrite2015nonequilibrium,chetrite2015variational}
We extend and apply an optimization algorithm that solves this variational expression and computes the optimal control force to sample rare dynamical phases.\cite{das2019variational} We show that this algorithm can be used to solve the inverse design problem, deciphering how rare fluctuations encode stability away from equilibrium.

\section{Variational design procedure}
We outline below an inverse design algorithm for the self-assembly of sheared DNA-coated colloids into different target nanoclusters.
The algorithm is based on a variational principle relating rare fluctuations in an ensemble of trajectories conditioned on evolving a target structure, to effective forces achieving the target as the typical dynamical state. Working with a trajectory ensemble, where the probability distribution is known, circumvents the difficulty of not knowing the distribution of configurations within a nonequilibrium steady-state. To solve the variational problem, we have used generalized response relations for the gradients of the steady-state trajectory probability to a change in the inter-particle and externally driven forces. 

\subsection{Model details}
For concreteness, we consider a model of \(N\) colloidal particles in a cubic box of length \(L\), evolving with an overdamped Langevin equation of the form,
\begin{equation}\label{eq:eom}
\gamma\dot{\mathbf{r}}_{i}=\mathbf{u}_{i}+\pmb{\eta}_{i}
\end{equation}
where \(\dot{\mathbf{r}}_{i}\) are time derivatives of the coordinates of the \(i\)-th particle and \(\mathbf{u}_{i}\) are the forces acting on it. The friction coefficient of the colloids with the thermal bath is denoted  \(\gamma\) and \(\pmb{\eta}_{i}\) are Gaussian white noise that satisfy
\begin{equation}
\langle\pmb{\eta}_{i}(t)\rangle=0 \,, \quad \langle\pmb{\eta}_{i}(t)\pmb{\eta}_{j}(t')\rangle=2\gamma k_{\mathrm{B}}T \mathbf{I}_{3}\delta_{ij} \delta(t-t')
\end{equation}
where \(\mathbf{I}_{3}\) is the \(3\times3\) identity matrix and $k_{\mathrm{B}}T$ is Boltzmann's constant times the temperature. The angular brackets denote an averaging operation over the random noise distribution. As we consider dynamics in the presence of a shear flow, we use Lees Edwards boundary conditions.\cite{lees1972computer} 

\begin{figure}[t]
\centering
\includegraphics[width=8.5cm]{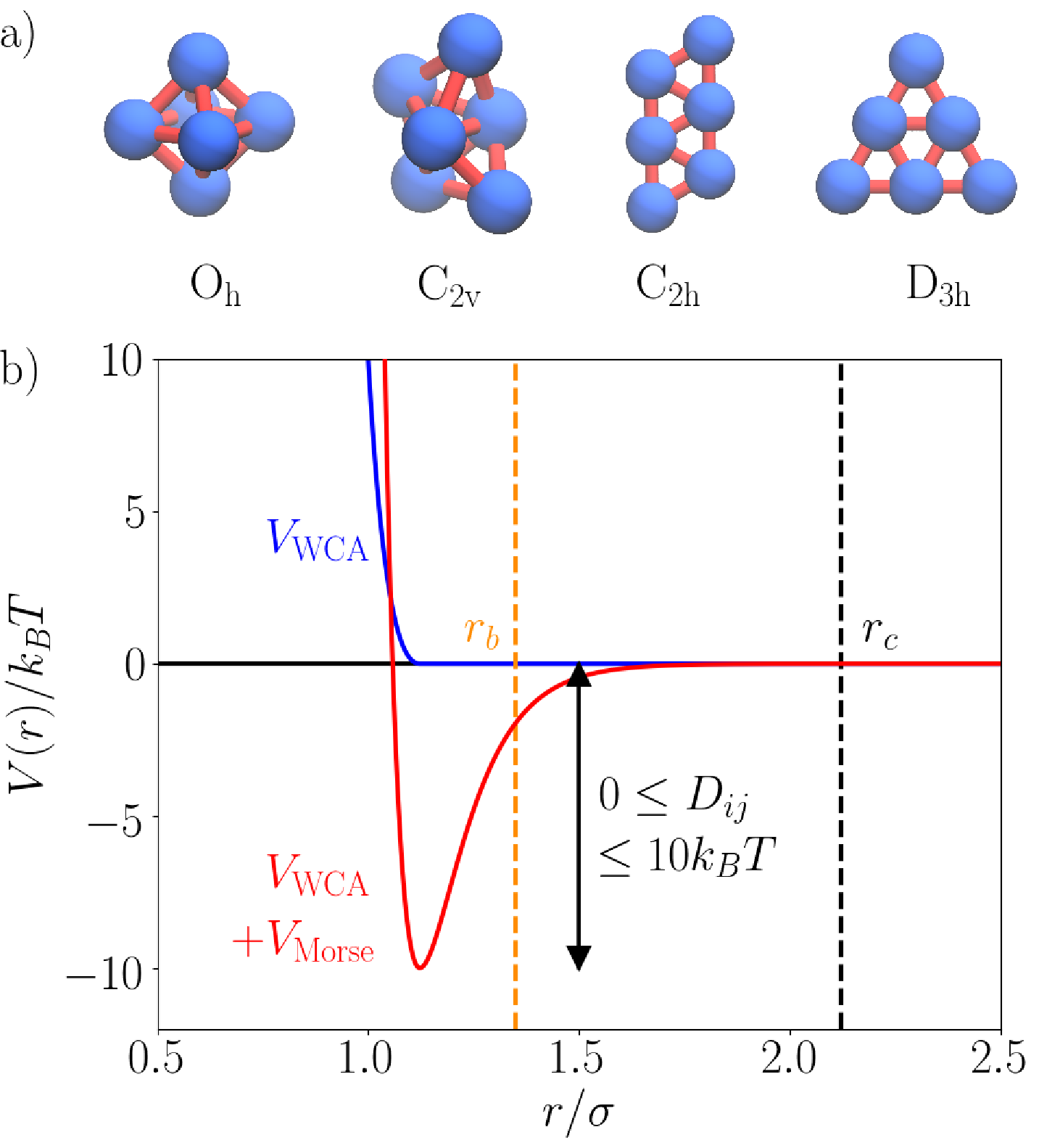}
\caption{Model details for self-assembly of DNA-labeled colloids. a) Examples of finite rigid and nonrigid nanoclusters for which we have studied design principles, along with the corresponding point groups for molecular symmetries denoted underneath. b) Graphical forms of the potential energy functions, the WCA potential (blue) and the WCA and Morse potential combined (red) for \(D_{ij}=10\kB T\). The orange dashed line denotes the bond cutoff \(r_{b}=1.35\sigma\) and the black dashed line denotes the potential cutoff \(r_{c}=2.12\sigma\).}
\label{fig1}
\end{figure}

We use an ansatz of DNA-labeled spherical isotropic colloids as programmable building blocks for self-assembly. The interaction between these colloids, mediated by the DNA molecules attached to their surface, consists of a volume-exclusion repulsion and a short-range attraction.\cite{rogers2011direct} The effective interaction strengths and the pairwise specificity can be independently tuned by varying the sequences of the grafted DNA molecules. During self-assembly, the short-range forces generate a competition between local and global order that leads to frustration and unique phase behavior and dynamical effects.\cite{rogers2015programming,manoharan2015colloidal} This system has been computationally and experimentally demonstrated to form finite nanoclusters with specific target structures.\cite{meng2010free,zeravcic2014size} The high-dimensional design space has the possibility to offer multiple pathways to stabilize any cluster out of the many nearly degenerate states formed without the specificity of the DNA-mediated attraction. To illustrate the performance of the variational algorithm, we consider the nonequilibrium self-assembly of 21 such rigid and nonrigid clusters, some examples of which are demonstrated in Figure \ref{fig1}a.

We examine the self-assembly of these colloidal particles under a constant linear shear flow. 
Shear flows are known phenomenologically to alter the stability of compact and extended colloidal structures.\cite{xue1990shear,cheng2012assembly} A recent paradigm of colloidal assembly being increasingly explored is that in a microfluidic device, where the confining walls generate a strong shear on the assembling clusters.\cite{dou2017review,nikoubashman2017self} This system offers a canonical nonequilibrium setting to explore inverse design principles.
Taken together, the forces acting on the \(i\)-th colloid are
\begin{equation}\label{eq:force}
\mathbf{u}_{i}=\mathbf{f}^{\mathrm{S}}_{i}(\mathbf{r}_{i})-\pmb{\nabla}_{i}\sum_{j\neq i} V(r_{ij})
\end{equation}
where $V(r)=V_{\mathrm{WCA}}(r)+V_{\mathrm{Morse}}(r)$ and \(V_{\mathrm{WCA}}(r_{ij})\) is a WCA pair potential representing the volume exclusion interactions, and \(V_{\mathrm{Morse}}(r_{\ij})\) denotes the DNA-mediated short-range pairwise attraction. The force due to a shear flow, \(\mathbf{f}^{\mathrm{S}}_{i}(\mathbf{r}_{i})\), has the form
\begin{equation}
\mathbf{f}^{\mathrm{S}}_{i}(\mathbf{r}_{i})=fz_{i}\mathbf{\hat{x}}
\end{equation}
which has magnitude $f$ and generates a constant gradient of the $x$ component of the velocity along the $z$ direction. 
The WCA pair potential has the functional form
\begin{align}
V_\mathrm{WCA}(r_{ij})&= 4\epsilon \left[\left( \frac{\sigma}{r_{ij}}\right )^{12}-\left (\frac{\sigma}{r_{ij}} \right)^{6} \right] +\epsilon \, , \quad r_{ij}<2^{1/6}\sigma \nonumber\\
&=0 \, , \quad  r_{ij}\geq2^{1/6}\sigma
\end{align}
with particle diameter $\sigma$ and energy scale $\epsilon$. The attractive Morse potential has the functional form
\begin{equation}\label{eq:morse}
V_{\mathrm{Morse}}(r_{ij})=D_{ij}\left( e^{-2\alpha(r_{ij}-2^{1/6}\sigma)}-2e^{-\alpha(r_{ij}-2^{1/6}\sigma)} \right)
\end{equation}
where $D_{ij}$ is the magnitude of the bond energy and $\alpha$ determines its width.  

We work in units of \(k_{\mathrm{B}}T=1\), \(\gamma=1\) and \(\sigma=1\). The natural time scale with these units is \(t_{0}=\gamma\sigma^{2}/\kB T\) and times are expressed in these units throughout. We set \(\epsilon=10k_{\mathrm{B}}T\) and \(\alpha^{-1}=\sigma/10\). The attractive energy scale \(D_{ij}\) and the shear flow rate \(f\) are tuned as variational parameters to induce self-assembly. They have been restricted to vary within the range \(0\leq D_{ij}\leq 10k_{\mathrm{B}}T\) and \(0\leq f\leq 50 k_{\mathrm{B}}T/\sigma^{2}\) to avoid large relaxation times and to stay within the overdamped regime. Figure~\ref{fig1}b shows the potentials for the inter-particle interactions. The Morse potential and its force have both been truncated and shifted, using the Shifted Forces approximation,\cite{toxvaerd2011communication} to decay smoothly to zero at \(r_{c}=2.12\sigma\). 

In order to avoid finite size effects in the formation of small clusters, we study a low packing fraction of  \(\phi=0.01\).  We use a first order Euler discretization for the equation of motion in Eq.~\ref{eq:eom}. Since the potentials in Eq.~\ref{eq:force} are narrow and short-range, we have to use a small timestep of \(5\times10^{-5}t_{0}\) in order to sample the potentials accurately. We have used trajectories of length ranging from \(\tau/t_{0}= 2.5\times10^{3}\) to \(10^{4}\). 

\subsection{Variational principle}
In order to uncover design principles for self-assembly, we consider the task of finding the set of forces that fulfill the condition of assembling a target structure as the typical state of the system in the long time limit.  Such tasks in stochastic dynamics are generalizations of Brownian bridges and known to have unique solutions.\cite{chetrite2015variational} They have played an important role recently in the application of large deviation theory to physical systems driven far from equilibrium.\cite{nemoto2016population, ray2018importance,ferre2018adaptive,grandpre2018current,das2019variational}

We start by defining an observable \(A_{\tau}\) as a time averaged indicator function for a target cluster,
\begin{equation}
A_{\tau}[\mathbf{r}^{N}(t)]=\frac{1}{\tau}\int_{0}^{\tau}\mathbf{1}[\mathbf{r}^{N}(t)]\;dt
\end{equation}
where $\mathbf{1}[\mathbf{r}^{N}(t)]=1$ for a configuration satisfying a geometric criterion consistent with a target cluster and $\mathbf{1}[\mathbf{r}^{N}(t)]=0$ otherwise, for each time $t$ along a trajectory $\mathbf{r}^{N}(t)$ of total duration \(\tau\). The average value of the observable quantifies the yield of the target cluster. For all colloidal clusters considered, $\mathbf{1}$  is computed by constructing a bond-connectivity matrix. A cutoff of \(r_{b}=1.35\sigma\) has been used to define a bond between two particles.  Indicator functions for rigid target clusters are then uniquely determined by permutation-invariant measures of this connectivity matrix.\cite{honeycutt1987molecular} For nonrigid target clusters, along with the bond-connectivity matrix, we additionally consider measures of the geometry of the cluster for defining the indicator function.

Rather than considering trajectories conditioned on a particular value of $A_{\tau}$ directly, which is numerically cumbersome, we work within an ensemble equivalent representation.\cite{chetrite2013nonequilibrium} Using a counting parameter \(\lambda\), we can statistically bias a system towards a particular value of $A_{\tau}$ within a nonequilibrium steady state. The cumulant generating function $\psi(\lambda)$ is the partition function associated with the trajectory ensemble under the statistical action of $\lambda$,
\begin{equation}
\psi(\lambda)=\lim_{\tau\to\infty}\frac{1}{\tau}\ln\left\langle e^{\lambda\tau A_{\tau}}\right\rangle _{0}
\end{equation}
where the angular brackets denote a path average over trajectory probability \(P_{0}[\mathbf{r}^{N}(t)]\), as
\begin{equation}
\left\langle e^{\lambda\tau A_{\tau}}\right\rangle _{0}=\int D[\mathbf{r}^{N}(t)]\exp\left( \lambda\tau A_{\tau}[\mathbf{r}^{N}(t)]\right) P_{0}[\mathbf{r}^{N}(t)]
\end{equation}
The subscript 0 refers to the average being computed in a reference ensemble where the particles do not typically show the desired self-assembly behavior. For this reference system we have chosen an equilibrium ensemble of colloids interacting only with the WCA repulsive forces, \textit{i.e.}, \(D_{ij}=f=0\), such that \(\mathbf{u}_{i}=-\pmb{\nabla}_{i}\sum_{j\neq i}V_{\mathrm{WCA}}(r_{ij})\) which is denoted as \(\mathbf{F}^{\mathrm{WCA}}_{i}(\mathbf{r}^{N})\). 

When the optimizable parameters are tuned to vary \(\mathbf{u}\), the trajectory probability \(P_{0}[\mathbf{r}^{N}(t)]\) changes to \(P_{\mathbf{u}}[\mathbf{r}^{N}(t)]\). The cumulant generating function can be estimated in the modified ensemble as,
\begin{align}\label{eq:cgf}
\psi(\lambda)&=\lim_{\tau\to\infty}\frac{1}{\tau}\ln \int D[\mathbf{r}^{N}(t)]e^{\lambda\tau A_{\tau}} \frac{P_{0}[\mathbf{r}^{N}(t)]}{P_{\mathbf{u}}[\mathbf{r}^{N}(t)]}P_{\mathbf{u}}[\mathbf{r}^{N}(t)] \nonumber\\
&=\lim_{\tau\to\infty}\frac{1}{\tau}\ln\left\langle e^{\lambda\tau A_{\tau}+\Delta S[\mathbf{u}]}\right\rangle _{\mathbf{u}}
\end{align}
where the functional form of the relative action \(\Delta S[\mathbf{u}]\) can be derived from Onsager-Machlup theory,\cite{onsager1953fluctuations}
\begin{align}
\Delta S[\mathbf{u}]&=S[\mathbf{u}]-S[\mathbf{F}^{\mathrm{WCA}}]\nonumber\\
&=\int_{0}^{\tau}\sum_{i=1}^{N}\frac{(\dot{\mathbf{r}}_{i}-\mathbf{u}_{i})^{2}-(\dot{\mathbf{r}}_{i}-\mathbf{F}_{i}^{\mathrm{WCA}})^{2}}{4\gamma\kB T}dt
\end{align}
with the integral being computed in the Ito sense. This change of measure analogous to a Girsanov transform\cite{chetrite2015nonequilibrium} relates the original likelihood of self-assembly in the reference ensemble to the ensemble under the control force. 

Since the exponential is a convex function, we apply Jensen's inequality to Eq. \ref{eq:cgf}
\begin{equation}
\psi(\lambda)\geq\lim_{\tau\to\infty}\frac{1}{\tau}\langle \lambda\tau A_{\tau}+\Delta S[\mathbf{u}]\rangle_{\mathbf{u}}
\end{equation}
 to obtain a variational expression for the cumulant generating function. In the long time limit for \(\tau\to\infty\), we can replace trajectory averages with static averages and simplify the relative action using the equation of motion for \(\dot{\mathbf{r}}_{i}\). Hence we arrive at our final variational expression,
\begin{equation}\label{eq:var}
\psi(\lambda)\geq\langle\Omega[\mathbf{u}]\rangle_{\mathbf{u}}=\left\langle \lambda\mathbf{1}(\mathbf{r}^{N})-\sum_{i=1}^{N}\frac{(\mathbf{u}_{i}-\mathbf{F}_{i}^{\mathrm{WCA}})^{2}}{4\gamma\kB T}\right\rangle _{\mathbf{u}}
\end{equation}
where $\langle\Omega[\mathbf{u}]\rangle_{\mathbf{u}}$ is the target function to optimize.

For a bounded observable like the indicator function, a large value of \(\lambda\) enforces the desired conditioning. The problem of saturating the variational inequality is known to have a unique solution when \(\mathbf{u}({\mathbf{r}^{N}})\) can take all possible functional forms, the optimal force being a generalization of Doob's h-transform.\cite{chetrite2015variational} Optimizing \(\langle\Omega[\mathbf{u}]\rangle_{\mathbf{u}}\) will lead to a set of many-body forces for assembling target clusters in high yield. In practice the use of only one-body and two-body forces in Eqs.~\ref{eq:force}-\ref{eq:morse} need not saturate the inequality. The second term in the variational expression is associated with the Kullback-Leibler divergence between the reference and conditioned trajectory ensembles. This term enforces the smallest excess force out of all possible control forces, and thus acts as a regularizer in the optimization process. While the solution to Eq.~\ref{eq:var} uniquely selects the force that shows fluctuations closest to rare fluctuations in the original ensemble, it is not a unique inverse design criterion and alternatives can in principle be constructed.\cite{rotskoff2017geometric} However, the optimization scheme that we construct in the next section can be generally extended to other functional forms of regularizers.

\subsection{Stochastic gradient descent}
To numerically optimize Eq.~\ref{eq:var}, we derive explicit gradients of the variational estimator \(\langle\Omega[\mathbf{u}]\rangle_{\mathbf{u}}\), using an algorithm that we have previously employed to estimate large deviation functions in nonequilibrium steady-states.\cite{das2019variational} The general form of the gradient with respect to any variational parameter \(c\in\{D_{ij},f\}\) is
\begin{align}\label{eq:grad}
\frac{\partial\langle\Omega[\mathbf{u}]\rangle_{\mathbf{u}}}{\partial c}&=\left\langle \frac{\partial\Omega[\mathbf{u}]}{\partial\mathbf{u}}\frac{\partial\mathbf{u}}{\partial c}\right\rangle _{\mathbf{u}} \nonumber\\
&-\int_{0}^{\infty}\left\langle \delta\Omega(t)\delta\left( \frac{\partial\dot{S}[\mathbf{u}]}{\partial\mathbf{u}}\frac{\partial\mathbf{u}}{\partial c}\right) (0)\right\rangle _{\mathbf{u}}dt
\end{align}
where \(\dot{S}[\mathbf{u}]\) is the time derivative of the action \(S[\mathbf{u}]\).
Equation \ref{eq:grad} is a generalized fluctuation-dissipation relation for a nonequilibrium response in the design parameter space. For computational purposes, we approximate the gradient expression by integrating the correlation function in the second term up to a fixed large time interval \(\Delta t=5t_{0}\). Due to the small density, we have to use a low variance estimate for the explicit gradient in Eq. \ref{eq:grad} for the specific case of optimizing the shear flow rate \(f\), see Appendix \ref{Sec:LowV}.

\begin{figure}[b]
\centering
\includegraphics[width=8.5cm]{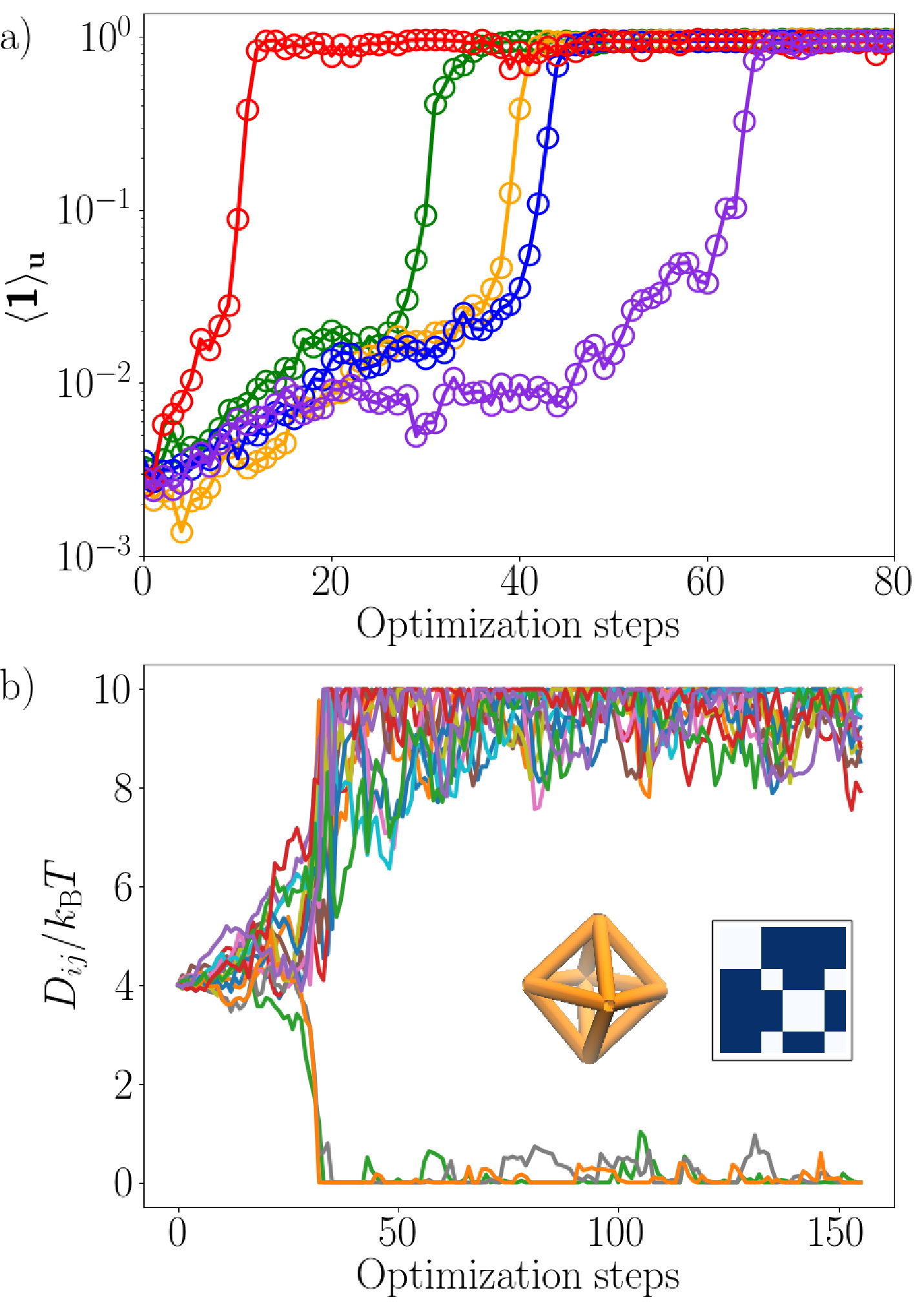}
\caption{Optimization procedure for an \(\mathrm{O_{h}}\) cluster. a) The convergence of the yield with increasing number of optimization steps. Different colors represent varying learning rates. b) Convergence of  \(D_{ij}\) for the green yield curve in (a). (\textit{Inset}) Bond structure and the corresponding MA  \(D_{ij}\) matrix for the \(\mathrm{O_{h}}\) cluster. Blue and white elements in the matrix denote bonds with \(D_{ij}=10\kB T\) and \(D_{ij}=0\), respectively.}
\label{fig2}
\end{figure}

The corresponding variational algorithm consists of a stochastic gradient descent optimization\cite{spall2005introduction} for a large value of \(\lambda\) in Eq. \ref{eq:var}. Starting from an initial point in parameter space \(\{D_{ij},f\}\), we simulate the dynamics of the system using Eq. \ref{eq:eom}, and after relaxation into a steady state, statistically estimate the explicit gradient of the variational estimator, Eq. \ref{eq:grad}. We perform stochastic gradient descent updating all variational parameters \(c\) at every step of the optimization, with the update rule at the \(n\)-th step being
\begin{equation}\label{eq:learning}
c_{n+1}=c_{n}+\alpha_{n}\frac{\partial\langle\Omega[\mathbf{u}]\rangle_{\mathbf{u}}}{\partial c}\Biggr\vert _{c_{n}}
\end{equation}
where the stochastic gradients are evaluated within a steady-state with the current value of the parameters, and \(\alpha_{n}\) is the learning rate for any of the $c$ parameters in the \(n\)-th optimization step. The level of noisy fluctuations in each parameter during the optimization process can be tuned independently through the corresponding learning rates. If the variational surface changes sharply, the learning rate has to be decreased with increasing \(n\) to anneal to the optimal solution basin. The learning rates have also been chosen individually for each example such that in each optimization step, the rate of change of \(D_{ij}/\kB T\) is in the range \([0.1,0.5]\) and that of \(f\sigma^{2}/\kB T\) is in the range \([1,5]\). 

 \subsection{Convergence and choice of $\lambda$}
To illustrate the performance of the optimization algorithm, we study the assembly of 6 particles into an octahedral (\(\mathrm{O}_{\mathrm{h}}\)) target cluster. An octahedron is not the highest yield cluster formed in a system of 6 hard sphere colloids with infinitely short-range attractions,\cite{meng2010free} and is formed in only 6\% yield with strong, nonspecific interactions.  Figure \ref{fig2}a shows the yield as a function of optimization steps with different learning rates and different trajectory noise histories. For all these examples, the yield is optimized over multiple orders of magnitude with the final converged value being close to 100\%. This change of the order parameter over several orders of magnitude arises from the observable being defined as the probability of forming the target cluster, and in cases where the change is more drastic, would necessitate the use of two different learning rates in Eq. \ref{eq:learning}. At a constant learning rate, the learning curves show two distinct regions, such that a gradual rise in yield is followed by a rapid convergence to the saturation value. 

Figure \ref{fig2}b shows the convergence of \(D_{ij}\) for one of the optimization runs. For 6 particles there are 15 distinct interactions, all of which are optimized. The starting point is a nonspecific attraction \(D_{ij}=4\kB T\) for all $ij$ pairs. The optimization curve shows two regions, an initial spreading of the \(D_{ij}\) values followed by a rapid permutation symmetry breaking and a clear segregation of the 15 interactions into 12 attractive and 3 repulsive parameters. The 12 attractive interactions are all statistically equal, as are the 3 repulsive interactions. The attractive interactions correspond to the 12 bonds in the connectivity matrix for the octahedron. The symmetry breaking is spontaneous and is aided by the initial noisy fluctuations during optimization. Different noise histories in the trajectory lead to a symmetry breaking for which different sets of \(D_{ij}\) parameters become attractive or repulsive. For the finite clusters considered, this symmetry breaking is general.
We refer to the specific \(D_{ij}\) solutions for the optimal yield of a target cluster as an \textit{alphabet}, and the particular \(D_{ij}\) in which there is a pairwise attractive interaction for every contact in the target structure as a \textit{Maximal Alphabet} (MA). This strategy has been previously shown to be effective in the self-assembly of short-range interacting colloids into small clusters. \cite{hormoz2011design,zeravcic2014size}
 
\begin{figure}[b]
\centering
\includegraphics[width=8.5cm]{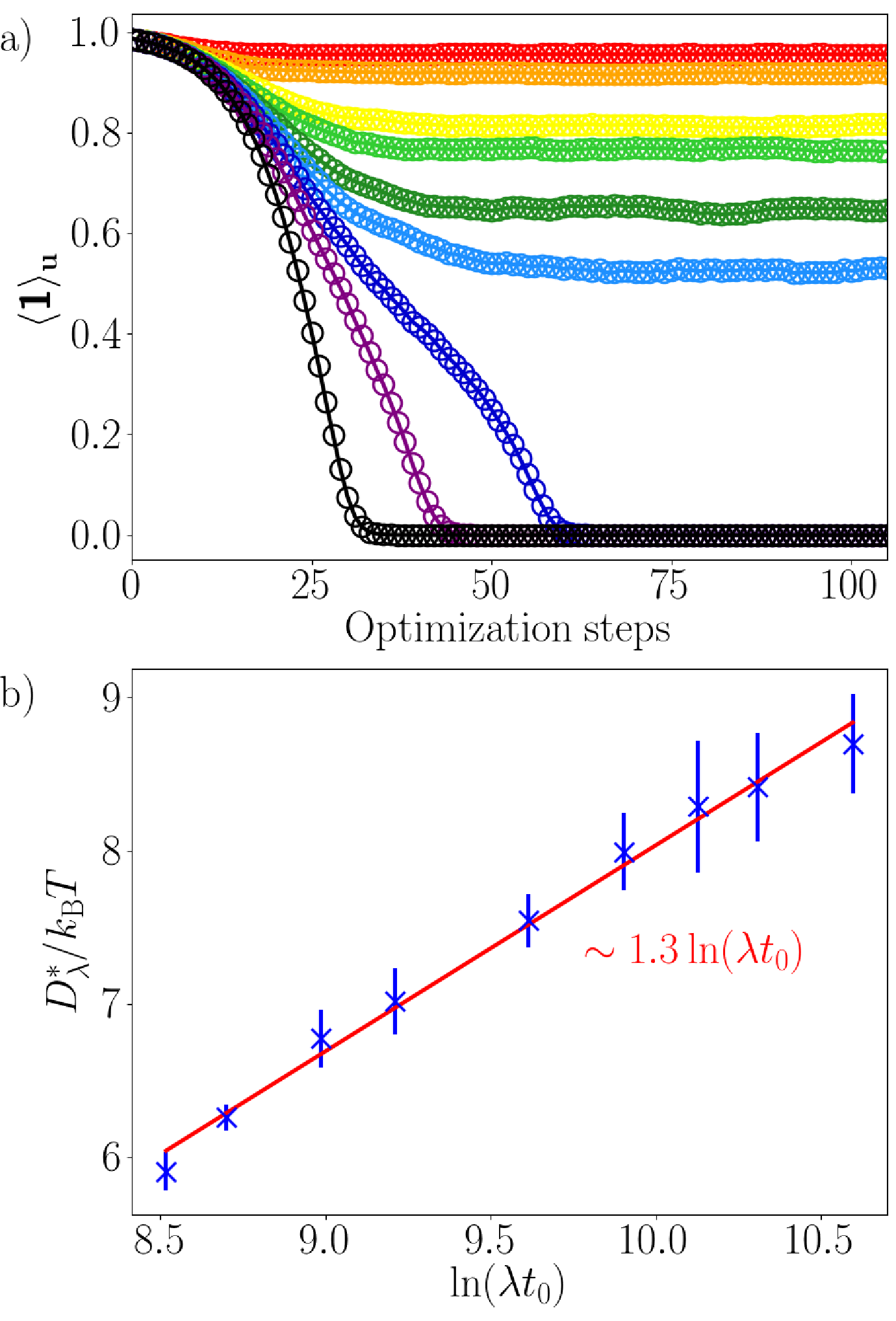}
\caption{Stability of the high-yield solution with $\lambda$. a) Convergence of the yield starting from the \(10\kB T\) MA solution. Different colors represent varying values of \(\lambda t_{0}\) in the range \([10^{3},4\times10^{4}]\), with the converged value of the yield increasing monotonically with \(\lambda\). b) Blue crosses with errorbars are the converged bond energy of the MA solution at varying values of \(\lambda\). Red line is a linear fit.}
\label{fig3}
\end{figure}

For the octahedral cluster, we have studied the stability of the MA solution for varying values of \(\lambda\). We find a bimodal structure of the variational surface, with the algorithm converging to either a MA solution or a  trivial solution \(\mathbf{u}=\mathbf{F}^{\mathrm{WCA}}\), depending on the value of \(\lambda\). 
Figure~\ref{fig3}a demonstrates the convergence of the octahedral yield with different values of \(\lambda t_{0}\) varying in the range \([10^{3},4\times 10^{4}]\), starting from an MA solution with \(10\kB T\) bond energies and a yield of 100\%. For moderate but decreasing values of \(\lambda\), the algorithm remains stable in the MA solution, but with monotonically decreasing yields and bond energies. Below a critical value of \(\lambda_{c}=5\times10^{3}/t_{0}\), the MA solution becomes unstable and the algorithm finds the \(\mathbf{u}=\mathbf{F}^{\mathrm{WCA}}\) solution. Rather than optimizing the yield in Eq. \ref{eq:var}, at small values of \(\lambda\) the second term is optimized. For some moderate \(\lambda\) values, the MA basin is only a local optimum and the crossover behavior shows \(\lambda\)-dependent hysteresis. We find this bistablity of the variational surface to be generic.
\begin{figure*}[t!]
\centering
\includegraphics[width=16cm]{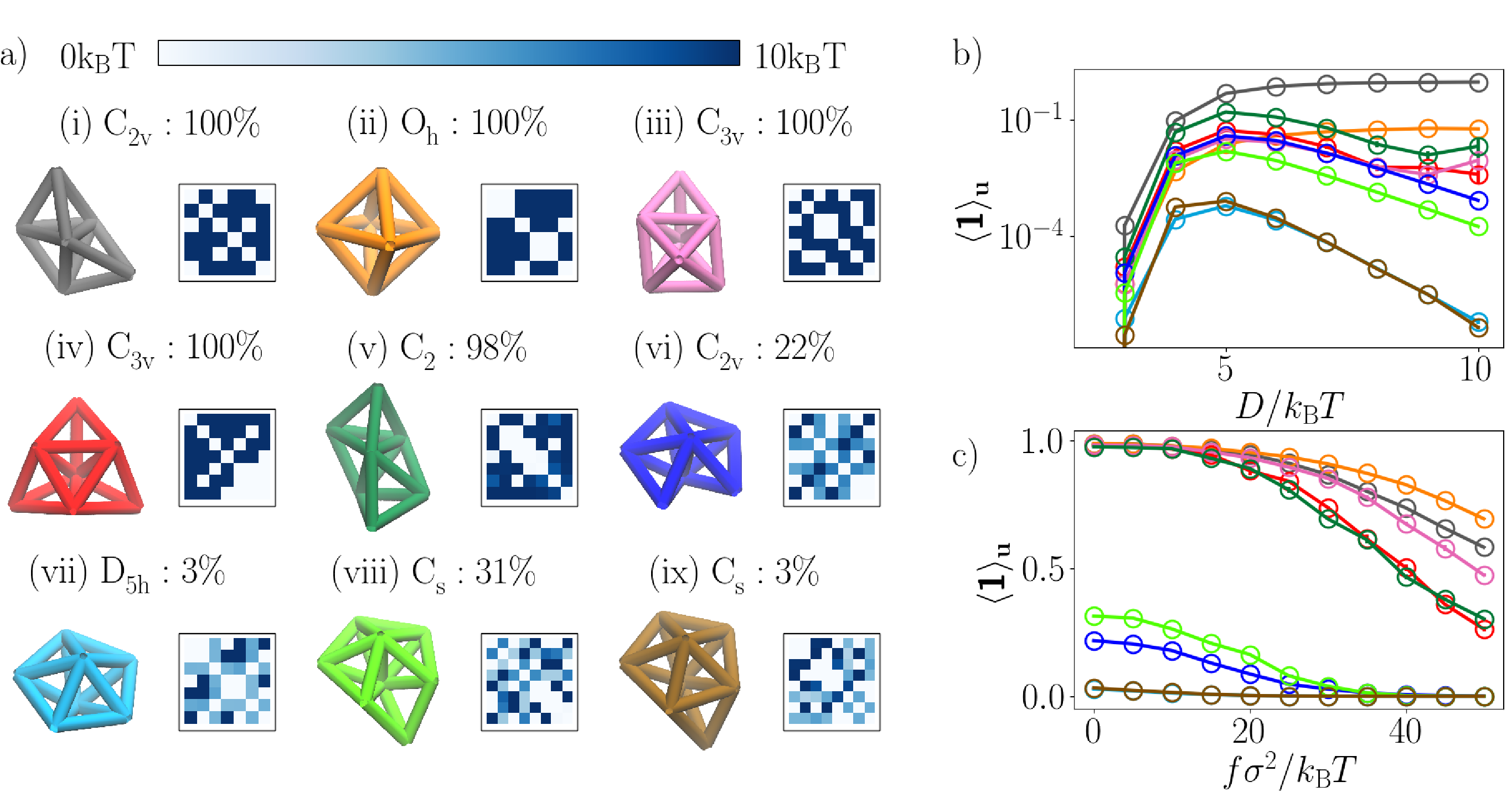}
\caption{Design principles for rigid clusters. a) Bond structure of clusters (i-ix), along with their corresponding point groups, optimal yields and their converged \textit{alphabets} on a color scheme indicated by the colorbar at the top. b) Yield as a function of a nonspecific attraction \(D_{ij}=D\) for fixed shear $f=0$. c) Yield as a function of the shear rate \(f\) for fixed optimal \textit{alphabets}. The colors of the points in b) and c) correspond to the colors of the clusters in a).}
\label{fig4}
\end{figure*}

Figure \ref{fig3}b shows the dependence of the average optimized bond energies of the converged MA solutions from the previous figure, as a function of varying \(\lambda\). This illustrates the depth of the MA basin in the parameter space. In the limit that each bond is formed independently, \(D_{\lambda}^{*}\) is expected to asymptotically vary as \(\sim 2 \kB T\ln(\lambda t_{0})\), as demonstrated in Appendix~\ref{App:Bias}. Over the range of $\lambda$ considered, we find a logarithmic dependence but with a different coefficient, \(\sim 1.3\ln(\lambda t_{0})\). This suggests that the free energy is approximately pairwise additive.

In the limit of large \(\lambda\), which in practice is 
chosen such that the estimate of the first term in Eq. \ref{eq:var} is at least an order of magnitude larger than the negative second term,
the variational algorithm can be used to automate the discovery of optimal forces for the self-assembly of clusters of arbitrary shapes and sizes, in a nonequilibrium steady-state. The optimal forces stabilize the target clusters in an arbitrary ensemble without accounting for the dynamics of transient relaxation towards its steady-state. For fixed number of tunable parameters, the computational cost scales linearly with system size since the only bottleneck is propagating a steady-state trajectory long enough to compute statistically converged gradients. The algorithm also scales linearly with the number of variational parameters, but with a small proportionality constant as all the gradients are estimated from the same trajectory. The use of the statistically estimated gradients significantly lowers the computational cost in contrast to numerically estimating the gradients from finite difference techniques by propagating multiple trajectories at different points in the parameter space. We next use our variational algorithm to study and rationalize the optimal design principles for a collection of rigid and nonrigid clusters.

\section{Design principles}
We have investigated the formation of small low-energy rigid and nonrigid clusters with 6,7 or 8 particles. We discover distinct design principles of these clusters and rationalize our findings by analyzing the response function of yield to the shear flow rate. We also demonstrate that the variational algorithm can obtain high yield optimal solutions even with constraints imposed on the total number of experimentally realizable design parameters. The design principles we obtain are expected to be general for the nonequilibrium self-assembly of short-range interacting colloids in a sheared steady-state.
\begin{figure*}[t]
\centering
\includegraphics[width=17cm]{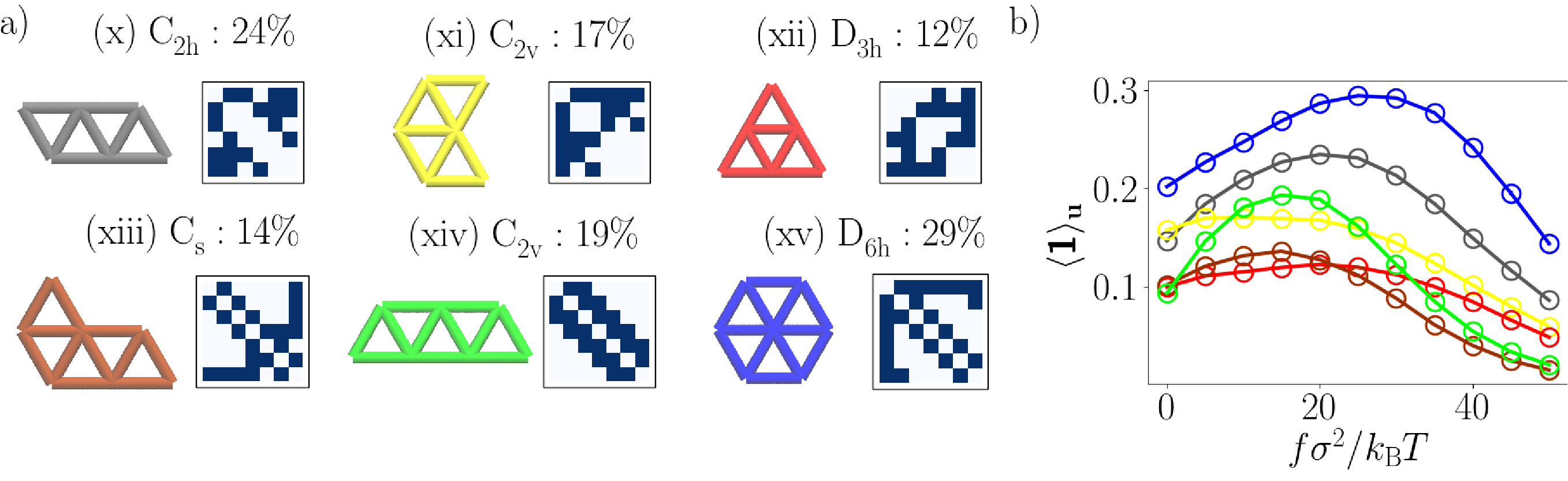}
\caption{Design principles for nonrigid clusters. a) Bond structure of clusters (x-xv), along with their corresponding point groups, optimal yields and their converged \textit{alphabets} on the same color scheme that was used in Fig. \ref{fig4}. b) Yield as a function of the shear rate for \(D_{ij}\) fixed at the optimal \textit{alphabet}. The colors of the points correspond to the clusters in a).}
\label{fig5}
\end{figure*}

\subsection{Rigid clusters}
We study the formation of a family of rigid clusters that are all known to be the lowest energy structures for systems of hard sphere colloids with identical infinitely short-range attractions. These clusters have previously been systematically enumerated and tabulated,\cite{arkus2009theoretical,arkus2009minimal} and their free-energy landscapes have been theoretically and experimentally studied.\cite{meng2010free} These finite clusters are the colloidal analogs to small molecules and have been shown to be involved in the controlled seeding and growth of polycrystalline phases and kinetically arrested gels.\cite{hecht2016kinetically,hermes2011nucleation} 

For each of these clusters, we optimize \(\{D_{ij},f\}\) to extremize the yield within this force ansatz. In the limit that the attractive interactions between the particles were infinitely short-range, there would be no internal low-energy distortion modes and the bond-connectivity matrix would correspond to a unique geometry. For our optimization, the indicator function refers to the corresponding bond-connectivity matrix conditions being satisfied. Figure. \ref{fig4}a summarizes the design principles discovered for these clusters. The point groups for the symmetries of each of these clusters have been indicated along with the highest yields obtained. For chiral \(\mathrm{C_{n}}\) clusters, the yields are the racemic yield. 

For each of these clusters, the corresponding optimal \textit{alphabet} discovered by the variational algorithm has also been indicated. We find that for clusters (i-iv), the optimal solution for \(D_{ij}\) is the MA. For the chiral \(\mathrm{C}_{\mathrm{2}}\) cluster (v), the optimal \textit{alphabet} is closely related to the MA but has a higher symmetry and is equivalent to a smaller \(3\times3\) \textit{alphabet}, while having the same yield. For clusters (vi-ix), all of which contain a radial 5-fold motif, the optimal yields are much less than 100\%, and the optimal \textit{alphabets} are not MA. The reason is the competition with structures with higher number of bonds. These lower energy structures would be geometrically unfeasible for infinitely short-range attractions, however in our model the short-range bonds have nonzero vibrations, which is sufficient to lead to the formation of the extra bonds. Unlike MA, the optimal \textit{alphabet} discovered by the variational algorithm penalizes the formation of these lower energy competing structures.

Figures \ref{fig4}b and \ref{fig4}c show two slices through the optimization landscape in the parameter space of \(\{D_{ij},f\}\). Figure \ref{fig4}b is a diagonal slice through \(D_{ij}\), such that all \(D_{ij}=D\), while fixing \(f=0\). We find that for the two 6 particle clusters (i) and (ii), there is a monotonic increase in yield with increasing \(D\). This suggests that even when the attractive forces are not infinitely short-range, both of these clusters are energetically the most stable configurations, and the only competing structures are higher in energy. When \(D=10\kB T\), the \(\mathrm{C}_{\mathrm{2v}}\) cluster (i) is formed with a yield of 93\% compared to the 6\% yield of the \(\mathrm{O}_{\mathrm{h}}\) cluster (ii), which is consistent with the stabilization due to the rotational entropy in the former.\cite{meng2010free} All the other clusters (iii-ix) in Fig. \ref{fig4}b show a turnover in yield with increasing \(D\). This 
is due to competing low energy structures that are formed at large enough nonspecific \(D\).
We expect this design principle of a turnover in yield with increasing nonspecific attractions to be general for larger clusters, since most larger clusters built from short-range interacting particles will contain the radial 5-fold motif.\cite{arkus2009minimal} The value of the attraction \(D\) at the yield turnover is determined by a competition between the energetic stabilization of the lower energy structure and the destabilization of structures with missing bonds.

Figure \ref{fig4}c shows a slice through the optimization landscape with varying the shear rate \(f\), while fixing \(D_{ij}\) at their optimal value found by the variational algorithm. We see that for this class of rigid clusters, the yield monotonically decreases with increasing shear rate. We expect this feature to be general for rigid clusters, since rigid clusters need no additional geometric stabilization that can be provided by shear, which only energetically destabilizes the bonds in the cluster. This perspective is confirmed in Sec.~\ref{Sec:Shear} using a linear response theory.

\subsection{Nonrigid clusters}

For our ansatz of short-range interacting colloids, clusters with \(N\) particles but fewer than \(3N-6\) bonds in total, and fewer than \(3\) bonds for every particle, are not minimally rigid in that they have zero energy deformation modes.\cite{arkus2009theoretical,arkus2009minimal} These clusters are not formed in high yield as stable ground state structures in equilibrium. We have used the variational algorithm to uncover optimal nonequilibrium design principles for a family of such nonrigid clusters. The clusters we have considered belong to a family of planar two-dimensional structures known as polyiamonds. They have been shown to self-assemble from colloids in the presence of a spatial heterogeneity, like in hydrodynamically driven assembly of sedimenting colloids in the presence of a substrate.\cite{pham2020two,niu2017self,perry2015two} Within our  control force ansatz, we investigate whether the shear flow planes are sufficient to stabilize these clusters.

For the optimization process, the indicator function for the cluster yield has been defined using both the bond connectivity matrix and the flatness of the clusters, as discussed in Appendix \ref{sec:Geo}. Figure~\ref{fig5}a shows the optimal design principles obtained for clusters (x-xiv). The variational algorithm converges on the MA solution for the \(D_{ij}\) parameters for all of these clusters. The yields, however, are not 100\% due to contribution from competing buckled configurations where the polyiamonds fold over the triangular faces to form tetrahedral motifs. Moreover, with the MA fixed, the optimal yield is at a non-zero shear flow. Figure \ref{fig5}b shows the yield as a function of the shear rate for fixed optimal alphabets. The location of the turnover in yield depends on the competition between geometric stabilization of the planar structure from the shear flow lines and energetic destabilization of the bonds. This design principle of planar two-dimensional clusters being stabilized by a shear flow appears to be general, and stands in contrast to the rigid clusters which are strictly destabilized by shear.

\subsection{Response of structure to shear}\label{Sec:Shear}

\begin{figure}[b!]
\centering
\includegraphics[width=8.5cm]{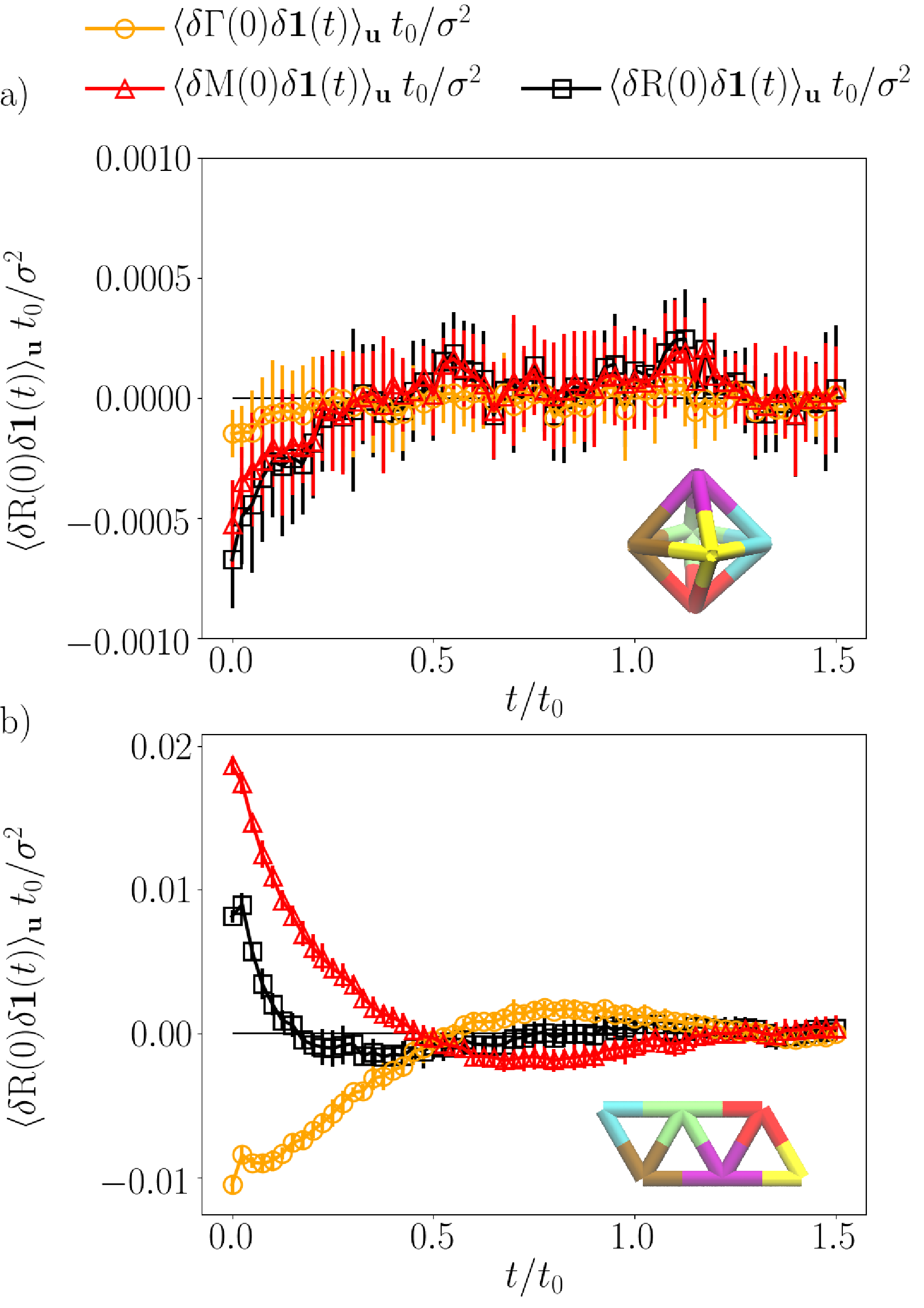}
\caption{Response of yield to shear for a rigid and a nonrigid cluster, with \(D_{ij}\) fixed at the corresponding optimal \textit{alphabets}, and shear fixed at \(f=5\kB T/\sigma^{2}\). a) Total correlation function (black squares) for an \(\mathrm{O_{h}}\) cluster (ii), and its torque (orange circles) and virial stress (red triangles) parts. b) The same correlation functions for the \(\mathrm{C_{2h}}\) cluster (x).}
\label{fig6}
\end{figure}

\begin{figure*}[th!]
\centering
\includegraphics[width=17cm]{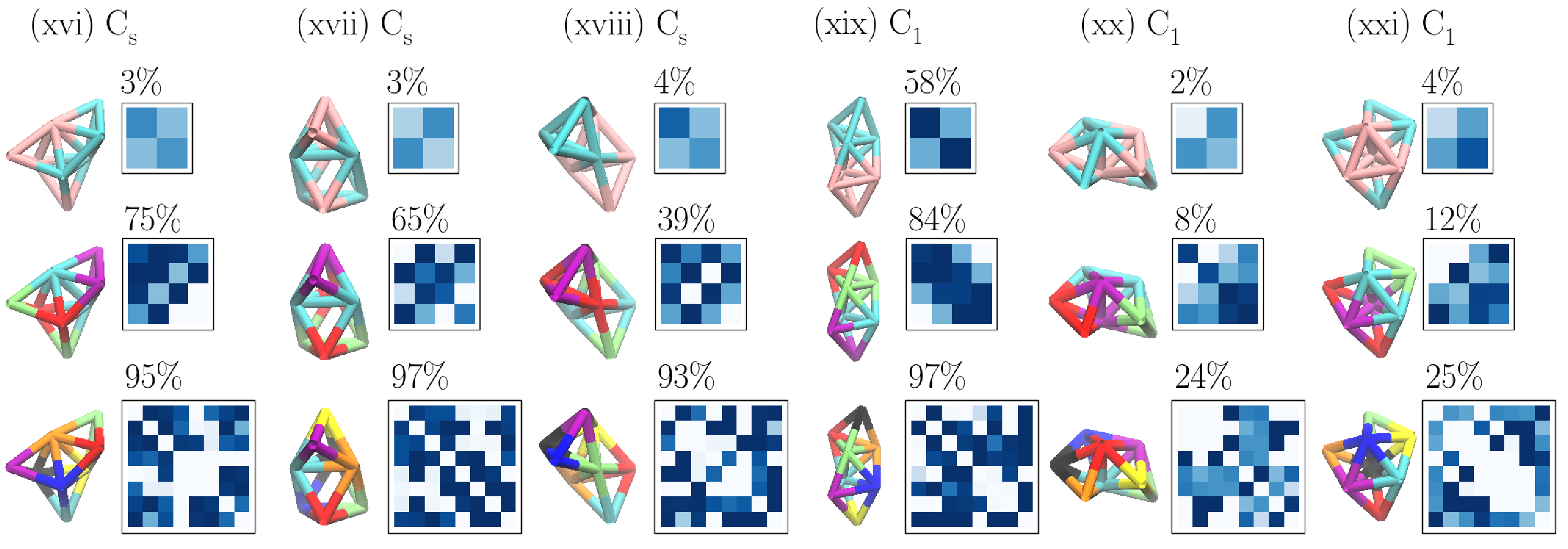}
\caption{Design principles for low symmetry clusters (xvi-xxi) with smaller alphabets. Each column corresponds to the point group of each cluster, its optimal yields and forces with 2$\times$2, 4$\times$4, and 8$\times$8 sized \textit{alphabets} in the three rows. The color scheme for the bonded structure refers to the optimal partitioning of 8 particles within 2, 4 or 8 labels, and the visualization of the \(D_{ij}\) matrices follow the same color scheme as in Fig. \ref{fig4}.}
\label{fig7}
\end{figure*}

The origin of the response of yield to shear flow is related to the relaxation dynamics of order parameter fluctuations in the unperturbed system. The response coefficients for rigid and nonrigid clusters can be understood using a generalized linear response theory.\cite{baiesi2009nonequilibrium,gao2019nonlinear} Keeping the \(D_{ij}\) parameters fixed, the linear response of the yield to a change in shear flow rate can be decomposed into two terms,
\begin{align}
\frac{\partial\langle\mathbf{1}\rangle_{\mathbf{u}}}{\partial f}&=\frac{1}{\kB T}\int_{0}^{\infty}dt\;
\left\langle \delta\Gamma(0)\delta\mathbf{1}(t)\right\rangle_{\mathbf{u}}\nonumber\\
&+\frac{1}{\kB T}\int_{0}^{\infty}dt\;\left\langle \delta\mathrm{M}(0)\delta\mathbf{1}(t)\right\rangle_{\mathbf{u}}
\end{align}
where we have used the low variance estimator for the correlation function  in Appendix~\ref{Sec:LowV}. Here, \(\Gamma\) is related to the dynamical torque acting on the cluster,
\begin{equation}
\Gamma=\frac{1}{2\gamma N}\sum_{i>j}[(\gamma\dot{x}_{i}-fz_{i})-(\gamma\dot{x}_{j}-fz_{j})](z_{i}-z_{j})
\end{equation}
and \(\mathrm{M}\) refers to the virial stress on the cluster due to internal forces,
\begin{equation}
\mathrm{M}=\frac{1}{2\gamma N}\sum_{i>j}[F_{i}^{x}-F_{j}^{x}](z_{i}-z_{j})
\end{equation}
where \(\mathbf{F}_{i}=-\pmb{\nabla}_{i}\sum_{j\neq i}\left[ V_{\mathrm{WCA}}(r_{ij})+V_{\mathrm{Morse}}(r_{ij})\right] \) is the conservative force acting on the \(i\)-th particle. \(\Gamma\) and \(M\) are time-reversal asymmetric and symmetric parts respectively\cite{speck2008role} of the full stochastic action gradient, \(R=\Gamma+M\). Decomposing in this form is necessary to preserve the deconvolution of the center-of-mass motion from the internal coordinates of the cluster.\cite{asheichyk2019response}

This linear response function is 0 at equilibrium due to the spatial parity symmetry of the system. Hence we have characterized the different components of this correlation function at a small value of shear \(f=5\kB T/\sigma^{2}\), for the rigid cluster (ii) and the nonrigid cluster (x), fixing \(D_{ij}\) to their corresponding MA interactions. The results are shown in Fig. \ref{fig6}. We find that the component coming from the virial stress has opposite signs at small times for the rigid and the nonrigid cluster, which accounts for the opposite signs of the gradient of the yield. The tumbling motion of the clusters in a shear flow couples positively with the internal order parameter in the case of a nonrigid cluster and leads to an increase in yield with increasing shear flow rate at small values of shear. The shear flow planes function as a spatial heterogeneity that is generally a precondition for stabilizing these planar clusters during self-assembly. At large shear, the yields of both rigid and nonrigid clusters are decreased with increasing shear due to the larger anti-correlation between the torque and the indicator function. 

These design principles in and out-of-equilibrium  are general in their scope of applicability for small clusters formed by DNA-coated colloids. Nevertheless, a key limitation of this approach is the linear system size scaling of the number of different kinds of DNA-labeled colloids required in order to assemble a cluster, evident in the corresponding quadratic scaling of the number of variational parameters. We have addressed this limitation in the next section.

\subsection{Smaller alphabets}

Engineering a system with an extensive number of specific interactions is difficult, even with DNA-coated colloids. It is advantageous in this regard to uncover \textit{alphabet}s that code for the minimal sufficient interactions to stabilize a target structure, in such a way that does not increase with increasing system size. For example, polymers and crystals are macroscopic structures that can be assembled with a finite number of specific interactions, as both only require a repeating microscopic number of components to be stabilized, either a sequence of monomers or a unit cell. For clusters that do not have a clear repeating unit, discovering optimal design principles with smaller \textit{alphabets} is nontrivial.

We have studied this problem by considering the 6 different low-symmetry 8-particle clusters (xvi-xxi)  shown in Fig. \ref{fig7}. For each of these clusters, there is no direct way to partition the interactions into 2 or 4 classes based on their bonding environment or symmetry. We have used the variational algorithm to optimize the yield of each of these clusters, with a 2 particle, 4 particle and 8 particle \textit{alphabet}, in which  \(D_{ij}\) has 3, 10 and 28 independent variational parameters respectively. Clusters (xvi-xix) have near 100\% yields for the full sized \textit{alphabet}. Clusters (xx) and (xxi) compete with higher bonded clusters containing the radial 5-fold motif, and so have an optimal yield of lower than 100\% even with an 8 particle \textit{alphabet}. Nevertheless, for all the clusters, a 4 particle \textit{alphabet} can give quite large yields in comparison to the maximum possible yield. The variational algorithm identifies the optimal way to partition the groups of interactions of these clusters despite the lack of clear symmetry. 

For the \(\mathrm{C}_{\mathrm{1}}\) cluster (xix), even a 2 particle \textit{alphabet} has a high yield, despite not having any exact two-fold symmetry. In this case, the algorithm has recognized a near-symmetry in the cluster and has partitioned it into 2 groups. The symmetry of these letters is close to the symmetrical alphabet that was discovered by the algorithm in a related \(\mathrm{C}_{\mathrm{2}}\) cluster (v) in Fig. \ref{fig4}. We expect this potential to discover optimal design principles for large clusters with a small number of groups, to be promising towards the self-assembly of experimentally realizable systems with practical constraints on the limits of bottom-up design.

\section{Discussion}
We have developed an inverse design algorithm for the self-assembly of colloidal clusters in a nonequilibrium steady-state. The formalism exploits a variational structure originating from large deviation techniques for importance sampling in trajectory ensembles. The algorithm optimizes the yield of clusters of arbitrary shapes, sizes and geometry by tuning control forces within an arbitrarily chosen ansatz, with statistically estimated explicit gradients. We have demonstrated the performance of the algorithm using an ansatz of DNA-labeled colloidal clusters self-assembled in a shear flow, and have obtained design rules for different families of rigid and nonrigid clusters. This algorithm scales linearly both in system size and in the number of optimizable parameters in the force ansatz, but its performance is  independent of the specific order parameter chosen for the self-assembly process. For example, the choice of a locally defined structural order parameter such as the density or the degree of crystallinity as the optimized observable can produce design principles for the assembly of extended dispersed or periodic structures out of equilibrium. Similarly, dynamical order parameters like the instantaneous flux between two stable states can also be optimized using the same variational procedure in a suitable trajectory ensemble. Hence this algorithm can be used to tune both structural and dynamical order parameters of clusters in a nonequilibrium steady-state to produce dynamical phases having no equilibrium analogs.

This variational algorithm differs from other available inverse design algorithms for soft matter in and out of equilibrium. The equivalent variational structure in configuration space for systems in thermal equilibrium, where the potential energy function is optimizable and explicit gradients can be statistically estimated by autodifferentiation, has been used extensively as the basis of both importance sampling and inverse design algorithms.\cite{valsson2014variational,bonati2019neural,sherman2020inverse}  This configuration space approach with explicit gradients is not feasible in nonequilibrium systems due to the probability measure being non-Boltzmann. Out of equilibrium, there have been theoretical approaches to rationalizing design principles in one or two component systems.\cite{nguyen2016design,whitelam2014self,whitelam2018strong} 

In the absence of a closed form expression for the configuration space measure, machine learning algorithms have been previously used to identity optimal design principles by tracking an order parameter during or at the end of finite-duration trajectories.\cite{long2014nonlinear,long2015machine,tang2016optimal,whitelam2020learning} Machine learning or neuroevolution based approaches are equivalent to estimating numerical gradients in the design space using finite-difference methods, and have similar convergence properties as explicit gradient based methods in the limit of small optimization steps.\cite{whitelam2020correspondence} However, since our multidimensional statistical gradient estimates are obtained using information from the same trajectory, our explicit gradient based method is expected to be advantageous in a high-dimensional design space as typically encountered in the bottom-up design of soft materials. 

Finally, a class of design algorithms employ a trajectory ensemble based approach to statistically estimate explicit gradients of the dynamical response in colloidal systems\cite{miskin2016turning,goodrich2020self} and are formally closest to our approach. These algorithms probe the transient dynamical response of self-assembly trajectories, which however do not predict the long-time properties of the self-assembled cluster in a nonequilibrium steady-state. In our algorithm, we have explicitly evaluated the long-time steady-state limit and arrived at the novel fluctuation-dissipation relation in design space in Eq. \ref{eq:grad}. This will enable our explicit-gradient based method to be directly used to optimize both structural and dynamic properties of driven phases of soft matter, and automate the discovery of new functional materials.
\vspace{0.5cm}

\noindent {\bf Acknowledgements} A.D. and D.T.L. were supported by the U.S. Department of Energy,  Office of Basic Energy Sciences through Award Number DE-SC0019375.
\vspace{0.5cm}

\noindent {\bf Data Availability} The data that support the findings of this study are openly available in Zenodo at \href {\doibase 10.5281/zenodo.4289235}{https://doi.org/10.5281/zenodo.4289235}.\cite{avishek_das_david_t_limmer_2020_4289235}

\appendix

\section{Estimating explicit gradients in the low-density limit}
\label{Sec:LowV}
Using Eq. \ref{eq:grad}, the second term in the explicit gradient with respect to the shear flow rate \(f\) is
\begin{align}
&\left\langle \delta\Omega(t)\delta\left( \frac{\partial\dot{S}[\mathbf{u}]}{\partial\mathbf{u}}\frac{\partial\mathbf{u}}{\partial f}\right) (0)\right\rangle _{\mathbf{u}}\nonumber\\
&=\frac{1}{2\gamma\kB T}\left\langle \delta\Omega(t)\delta\left( \sum_{i=1}^{N}\eta_{i}^{x}z_{i}\right) (0)\right\rangle_{\mathbf{u}} 
\end{align}
where we have simplified the stochastic action using the equation of motion. Since \(z_{i}\) appears in the expression independently for each particle, and the density of the particles is vanishingly small, the \(z\)-diffusion timescale of the cluster diverges, and the correlation function takes a long time to converge. Thus any gradient estimate we obtain by integrating the correlation function to a finite time \(\Delta t\) will contain a systematic error. In order to obtain an unbiased gradient, we recognize that in the large \(\lambda\) limit we are working in, the major part of \(\Omega(t)\) is from \(\mathbf{1}(t)\), which by our definition depends only on the internal coordinates of the particle, and due to the spatial translation symmetry in our system, is decoupled from the center-of-mass diffusion. This decoupling is directly expressed by a regrouping of terms in the sum over particles,
\begin{align}
&\lambda\left\langle \delta\mathbf{1}(t)\delta \left(\sum_{i=1}^{N}\eta_{i}^{x}z_{i}\right) (0)\right\rangle _{\mathbf{u}}\nonumber\\
=&\frac{\lambda}{N}\left\langle \delta\mathbf{1}(t)\delta\left( \left[ \sum_{i=1}^{N}\eta_{i}^{x}\right] \left[ \sum_{i=1}^{N}z_{i}\right] \right) (0)\right\rangle _{\mathbf{u}}\nonumber\\
+&\frac{\lambda}{N}\left\langle \delta\mathbf{1}(t)\delta\left(  \sum_{\substack{i,j=1\\i>j}}^{N}(\eta_{i}^{x}-\eta_{j}^{x})(z_{i}-z_{j})\right) (0)\right\rangle _{\mathbf{u}}
\end{align}
where in the first term the \(z\) coordinate of the center of mass has been explicitly factored out. We identify that the indicator function does not correlate with the center-of-mass motion and so the first term is 0. We use only the second term to approximately evaluate the gradients of Eq. \ref{eq:var} with respect to the shear flow rate \(f\).

\section{Asymptotic dependence of solution on $\lambda$}
\label{App:Bias}
We analytically solve for the asymptotic dependence of the bond energy \(D\) as a function of the bias \(\lambda\), for the formation of one bond, independent from the dynamics of the other bonds. We can represent this simplified system as an equilibrium, two state Markov model with the rates of transition between the bonded and unbonded states \(k_{b}\) and \(k_{u}\), respectively. These rates are determined by their mean \(\nu=(k_{u}+k_{b})/2\) and their ratio \(k_{u}/k_{b}=\exp(\Delta F/\kB T)\), where  $\Delta F$  is  the free energy difference of the two states.

The cumulant generating function associated with fluctuations in the bonded state satifies a eigenvalue equation of the form,
\begin{equation}
W_\lambda r_\lambda = \psi(\lambda) r_\lambda
\end{equation}
where $r_\lambda$ is a right eigenvecture, and operator $W_\lambda$ is given by
\begin{equation}
W_\lambda = \begin{pmatrix}-k_{u}+\lambda & k_{u}\\k_{b}& -k_{b}\end{pmatrix}
\end{equation}
which is equal to the adjoint of the transition rate matrix when $\lambda=0$. For the optimal rates that generate the statistics equivalent to this rate matrix, we need the Doob's transform of the matrix.\cite{chetrite2015nonequilibrium} 
For this purpose we diagonalize the matrix to find the eigenvector corresponding to the dominant eigenvalue as \((r_{1},1)\) where
\begin{align}\label{eq:egnvec}
&r_{1}= \frac{e^{-\frac{\Delta F}{\kB T}}}{4\nu}\biggl[ (\lambda+2\nu)+e^{\frac{\Delta F}{\kB T}}(\lambda-2\nu)\nonumber\\
&-\sqrt{e^{2\frac{\Delta F}{\kB T}}(\lambda-2\nu)^{2}+(\lambda+2\nu)^{2}+2e^{\frac{\Delta F}{\kB T}}(\lambda^{2}+4\nu^{2})}\biggr]
\end{align}
and \(k_{u}\) and \(k_{b}\) have been rewritten with \(\Delta F\) and \(\nu\).

The modified rates that generate the optimal dynamics are given by \(\tilde{k}_{u}=k_{u}/r_{1}t_{0}\) and \(\tilde{k}_{b}=k_{b}r_{1}t_{0}\). The modified free energy difference corresponding to these rates is \(\Delta\tilde{F}=\kB T\ln(\tilde{k}_{u}/\tilde{k}_{b})\). Using Eq. (\ref{eq:egnvec}), in the \(\lambda\to\infty\) limit the optimal free energy goes as
\begin{equation}
\Delta\tilde{F}\sim\Delta F-2\kB T\ln(\lambda t_{0})
\end{equation}
In this limit, the free energy is dominated by the negative of the bond-formation energy \(D\), and hence the latter is asymptotically \(D\sim 2\kB T\ln(\lambda t_{0})\).
\begin{figure}[t!]
\centering
\includegraphics[width=8.5cm]{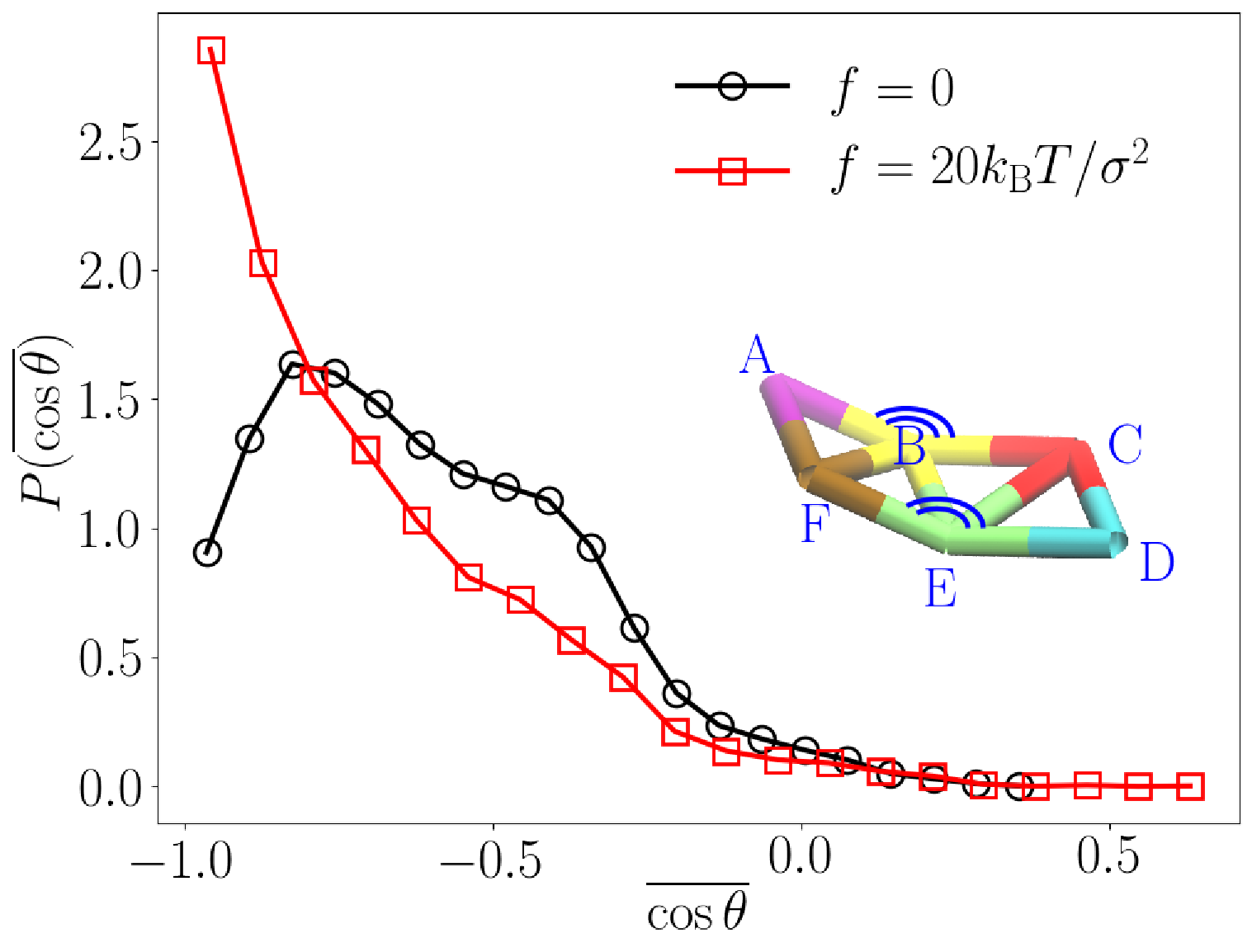}
\caption{Flatness distribution for the geometry of the \(\mathrm{C_{2h}}\) cluster (x), keeping the \(D_{ij}\) fixed at the optimal solution and changing the shear rate \(f\). (\textit{Inset}) The flatness is measured through the angles $\theta_{ABC}$ and $\theta_{DEF}$.}
\label{fig8}
\end{figure}

\section{Effect of shear on geometry of nonrigid clusters}
\label{sec:Geo}
Shear flow enhances the yield of nonrigid clusters (x-xv) by stabilizing a planar geometry and suppressing buckling modes. Here we have computed the probability distribution \(P(\overline{\cos\theta})\) of the average flatness, defined as \(\overline{\cos\theta}=[ \cos (\theta_{ABC})+ \cos (\theta_{DEF})]/2\), conditioned on the correct bond connectivity matrix for the cluster being satisfied. 
The angles, $\theta_{ABC}$ and  $\theta_{DEF}$ are defined in Fig. \ref{fig8}.
The flatness is \(-1\) for a perfectly planar geometry, but increases due to buckling and bending of the nonrigid cluster. For defining the indicator function for the nonrigid cluster, we used a flatness cutoff of \(\overline{\cos\theta}\le-0.8\). 
 We have looked at a population where the bond-connectivity matrix condition is satisfied but the flatness is unconditioned. This is illustrated in Fig. \ref{fig8} for the \(\mathrm{C_{2h}}\) cluster (x). We have kept the \(D_{ij}\) forces fixed at the optimized \textit{alphabet}, and plotted the distribution of flatness at two values of shear, at equilibrium with \(f=0\) and also at \(f=20\kB T/\sigma^{2}\) which is close to the optimal value for highest yield. We find that the planar geometry is a transient state at equilibrium, with the most probable states corresponding to the buckling of one or both of the angles $\theta_{ABC}$ and $\theta_{DEF}$. The shear flow destabilizes the buckled conformations and stabilizes the planar state instead, so that at \(f=20\kB T/\sigma^{2}\) the most probable conformation is the correct planar geometry.

\section*{References}

\end{document}